# Figures

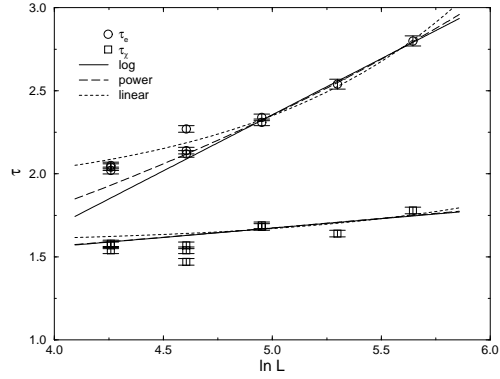

Figure 1: *The autocorrelation times of energy, $\tau_e$, and susceptibility, $\tau_\chi$ at the simulation point $K_0 = 0.263$, together with various types of fits to the data of the three largest lattices as discussed in the text.*

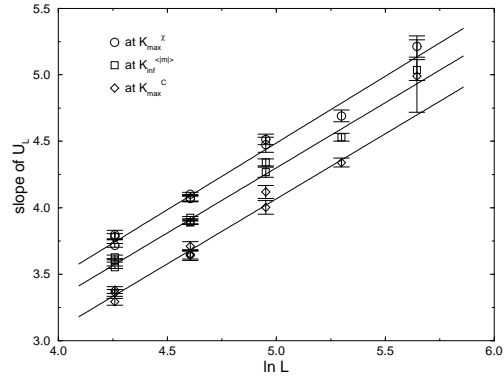

Figure 2: *Finite-size scaling of the Binder parameter slopes at sequences of $K(L)$-values for which the scaling variable $x \equiv (K(L) - K_c)L^{1/\nu}$ is constant. From the inverse slope of the least-squares fits we obtain estimates of $\nu = 0.996(28)$ ($K(L) = K_{\max}^{\chi'}(L)$), $\nu = 1.020(20)$ ($K(L) = K_{\inf}^{\langle |m| \rangle}(L)$), and $\nu = 1.021(33)$ ($K(L) = K_{\max}^C(L)$), respectively.*

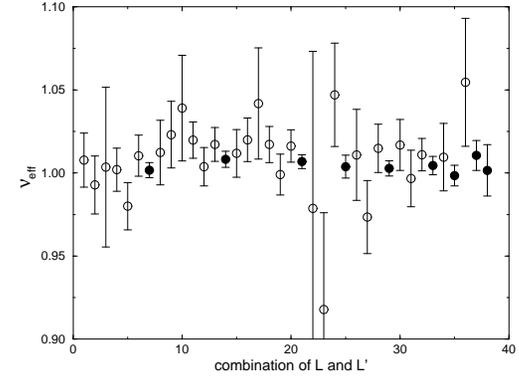

Figure 3: *Results for the effective exponents $\nu_{\text{eff}}$ defined in eq. (16). The x-axis labels the 38 possible combinations of lattices of size $N$ and $N'$, starting with $5\,000/10\,000$, $5\,000/20\,000$, and so on. The filled circles show all possible values for $N' = 80\,000$.*

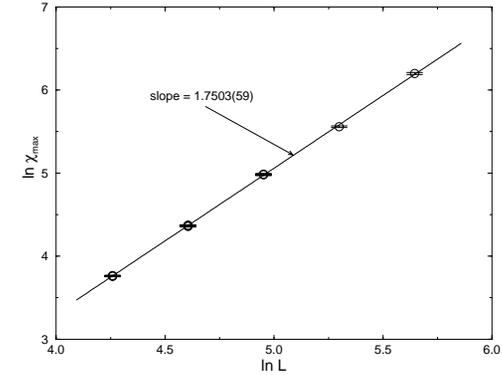

Figure 4: *Finite-size scaling of the (finite lattice) susceptibility maxima $\chi'_{\max}$. The slope of the linear least-squares fit yields $\gamma/\nu = 1.7503(59)$.*

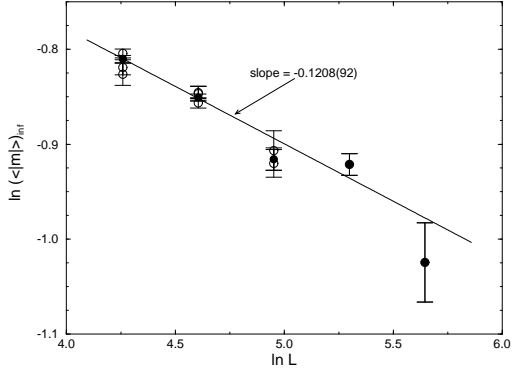

Figure 5: *Finite-size scaling of the (finite lattice) magnetization $\langle|m|\rangle$ at its point of inflection. The filled circles show the replica averages. From the slope of the linear least-squares fit we obtain $\beta/\nu = 0.1208(92)$.*

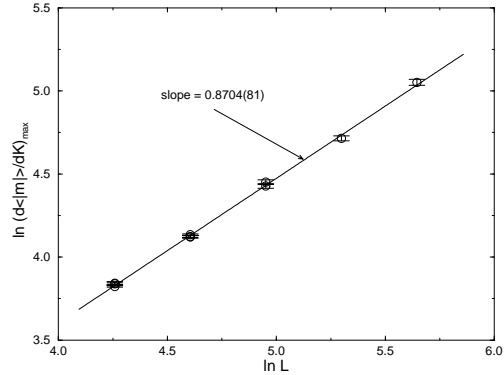

Figure 6: *Finite-size scaling of the maxima of $d\langle|m|\rangle/dK$, used to determine the points of inflection of $\langle|m|\rangle$. The slope of the linear least-squares fit is an estimate for $(1-\beta)/\nu = 0.8704(81)$.*

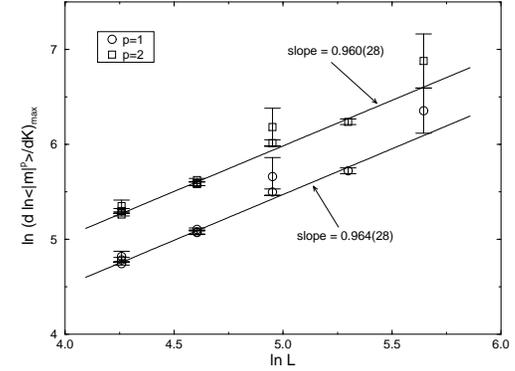

Figure 7: *Finite-size scaling of the maxima of the logarithmic derivatives $d\ln\langle|m|\rangle/dk$ $(p=1)$ and $d\ln\langle m^2\rangle/dk$ $(p=2)$. The slopes of the linear least-squares fits are estimates of $1/\nu$, resulting in $\nu = 1.037(31)$ $(p=1)$ and $\nu = 1.042(30)$ $(p=2)$, respectively.*

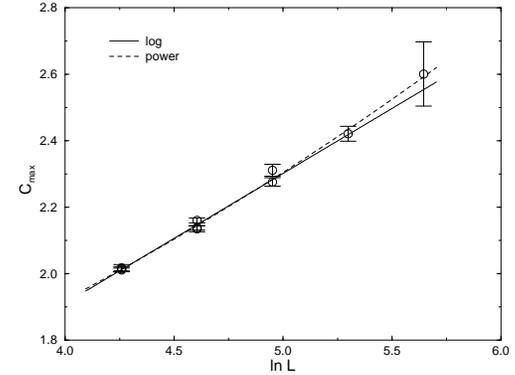

Figure 8: *Finite-size scaling of the specific-heat maxima $C_{\max}$. Also shown are least-squares fits to a logarithmic Ansatz, $C_{\max} = B_0 + B_1 \ln L$ (with $B_0 = 0.346(52)$, $B_1 = 0.391(12)$), and to a pure pure power-law Ansatz, $C_{\max} = cL^{\alpha/\nu}$ (with $c = 0.926(22)$, $\alpha/\nu = 0.1824(53)$).*

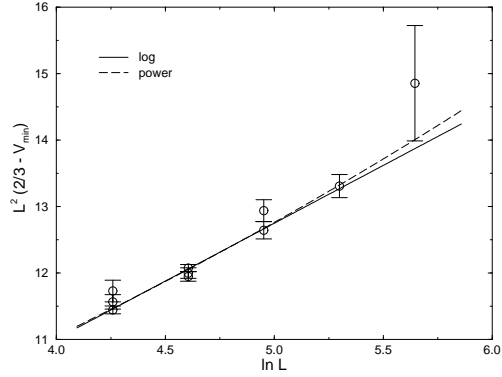

Fig. 9

Figure 9: *Finite-size scaling of the minima $V_{\min}$ of the energetic fourth-order parameter $V_L$. The solid and dashed lines show logarithmic and power-law fits, which are closely related to the corresponding fits to the specific-heat data in Fig. 8 (see text).*

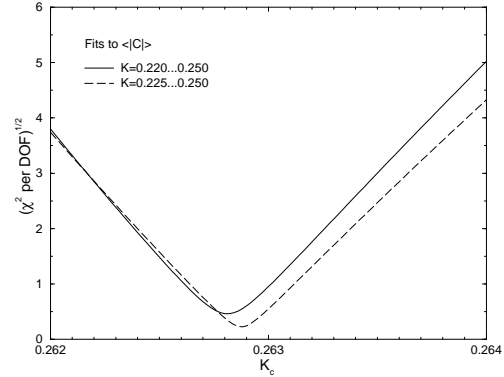

Fig. 10

Figure 10: *The chi-squared per degree of freedom ($\chi^2$ per $DOF$) of least-squares fits with fixed critical coupling $K_c$ to measurements of the improved susceptibility, $\chi_{\mathrm{imp}}/K = \langle |C| \rangle$, in the disordered phase. The minima are the best estimates for $K_c$ from this data.*

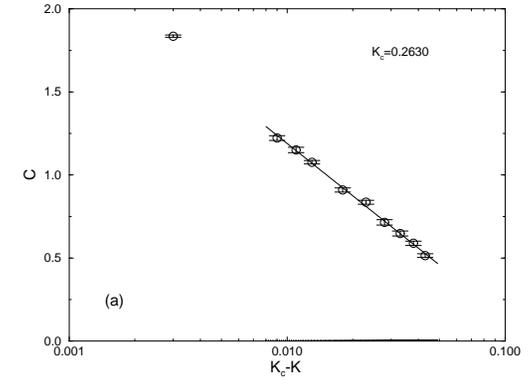

Fig. 11(a)

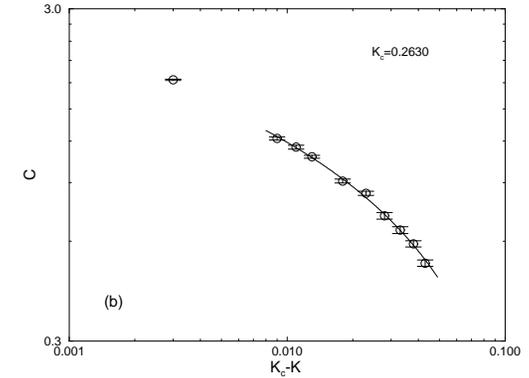

Fig. 11(b)

Figure 11: *The specific heat in the disordered phase near the critical coupling $K_c$. The semi-log plot in (a) demonstrates the consistency of our data with a logarithmic scaling behavior. The solid straight line shows a corresponding least-squares fit. In (b) the data and fit displayed in (a) are replotted in a log-log representation. Here a straight line would correspond to a power-law behavior.*

# Single-Cluster Monte Carlo Study of the Ising Model on Two-Dimensional Random Lattices

*Wolfhard Janke*[1], *Mohammad Katoot*[2] and *Ramon Villanova*[3]

[1] Institut für Physik, Johannes Gutenberg-Universität Mainz
55099 Mainz, Germany

[2] Department of Physical Sciences, Embry-Riddle Aeronautical University
Daytona Beach, Florida 32114, USA

[3] Grup de Física Teòrica and IFAE, Facultat de Ciències, Universitat Autònoma de Barcelona
08193 Bellaterra, Spain

**Abstract**

We use the single-cluster Monte Carlo update algorithm to simulate the Ising model on two-dimensional Poissonian random lattices with up to 80 000 sites which are linked together according to the Voronoi/Delaunay prescription. In one set of simulations we use reweighting techniques and finite-size scaling analysis to investigate the critical properties of the model in the very vicinity of the phase transition. In the other set of simulations we study the approach to criticality in the disordered phase, making use of improved estimators for measurements. From both sets of simulations we obtain clear evidence that the critical exponents agree with the exactly known exponents for regular lattices, i.e., that (lattice) universality holds for the two-dimensional Ising model.

# I. Introduction

Physically the concept of random lattices plays an important role in an idealized description of the statistical geometry of random packings of particles [1–3]. A prominent example is the crystallization process in liquids, and many statistical properties of random lattices have been studied in this context [4]. From a more technical point of view, random lattices provide a convenient tool to discretize space without introducing any kind of anisotropy [5]. In the past few years this desirable property of random lattices has been exploited in a great variety of fields. The applications range from quantum field theory or quantum gravity [5–7], the statistical mechanics of strings or membranes [8], to the solution of Laplace's equation in the context of diffusion limited aggregation [9], or the study of growth models for sandpiles [10], to mention a few. The preserved rotational, or more generally Poincaré, invariance suggests that field theories or spin systems defined on random lattices should reach the continuum or infinite volume limit faster than on regular lattices. An implicit assumption in this approach is that the concept of (lattice) universality, which is known to be true for spin systems on different *regular* lattices, carries over to random lattices. By this, one means that a system defined on different lattice discretizations should exhibit the same qualitative behavior once the physical length scale is much larger than the average lattice spacing. Even though this assumption appears very natural, previous numerical work [11, 12] on random lattices could only give weak evidence that it applies in this case as well.

In fact, in view of the equivalence of a random lattice system to a regular lattice system with impurity bonds derived a long time ago [13], the universality assumption might appear less trivial than naively expected [14]. In particular for the two-dimensional Ising model, according to the Harris criterion [15], random disorder is marginally important since the critical exponent of the specific heat is $\alpha = 0$. It should be noted, however, that in the random lattice case the equivalent distribution of impurity bonds exhibits complicated correlations, which makes the theoretical analysis even more subtle.

To investigate this point numerically, Espriu *et al.* [12] performed Monte Carlo (MC) simulations of the Ising model on a two-dimensional Poissonian random lattice with $N = 10\,000$ sites, linked together according to the Voronoi/Delaunay prescription [1–4]. Analyzing their data in the high- and low-temperature phase they obtained weak evidence that the critical exponents for the random lattice system agree with the regular Onsager values [16], which we have summarized for the reader's convenience in Table 1. Using a local Metropolis update algorithm as in Ref.[12], it would be very time consuming to obtain a significant improvement, especially in the vicinity of the phase transition where critical slowing down is a severe problem [17]. In the meantime much more efficient update algorithms have been discovered [18–20] which overcome this problem. Together with improved methods of data analysis [21] this now allows to simulate the model also at criticality with high precision and to study its finite-size scaling (FSS) behavior [22].

Table 1: *Exact critical exponents for the regular 2D Ising model.*

| $\alpha$ | $\nu$ | $\beta$ | $\gamma$ | $\eta$ | $\delta$ |
|---|---|---|---|---|---|
| 0(log) | 1 | 0.125 | 1.75 | 0.25 | 15 |

In this paper, we thus present and analyze two sets of extensive simulations of the Ising model on two-dimensional Poissonian random lattices of Voronoi/Delaunay type, varying in size from 5 000 to 80 000 sites. In the first set of simulations we concentrate on the very vicinity of the transition point and apply FSS techniques [22] to extract the critical coupling $K_c \equiv 1/k_B T_c$, and the exponents $\nu$, $\gamma/\nu$, $\beta/\nu$, and $\alpha/\nu$ [23]. The second set of data consists of simulations in the disordered phase for a random lattice of size 40 000 sites. Here we focus on the approach to criticality of the susceptibility and the specific heat, which yield independent estimates of $K_c$, and of the critical exponents $\gamma$ and $\alpha$.

To achieve the desired accuracy of the data in reasonable computer time we have applied the single-cluster algorithm [19] to update the spins. In the FSS region we further made extensively use of the reweighting technique [21], and in the disordered phase we took advantage of the fact that the average cluster size is an improved estimator for the susceptibility.

The paper is divided as follows. In Sec. II, we briefly describe the numerical construction of the random lattice and give a few simulation details. In Sec. III, we present a finite-size scaling analysis of our simulations near the transition point. Section IV is devoted to a discussion of our results in the disordered phase, and in Sec. V we close with a brief summary and a few concluding remarks.

## II. The model and simulation techniques

**A. Random lattice construction and properties:** In constructing the random lattices we followed closely the method described by Friedberg and Ren [7]. At first we draw $N$ random sites distributed uniformly in a unit square, thereby generating a so-called Poissonian distribution. For alternative distributions discussed in the literature see, e.g., Refs.[9, 24]. To link these sites according to the Voronoi/Delaunay prescription, we start by picking one site at random, locate its nearest neighbor and store this link along with its direction. Next a third site is searched for in a counter clock-wise sense by drawing a family of circles that pass through the first two sites and are centered on their bisector. Once the first triangle is completed this procedure continues with the same steps until all sites are linked. Some care must be exercised when approaching the boundaries of the lattice to ensure the periodic boundary conditions. To implement the nearest-neighbor search efficiently we subdivided the unit square into smaller boxes. The optimal box size is determined by two conflicting requirements. On the one hand, the box size should be large enough to ensure that nearest neighbors will be located with high probability in the same box or at least in the eight surrounding boxes. On the other hand, to minimize the time needed for testing all sites in a box, the box size should be as small as possible. We only performed a "trial and error" optimization based on heuristic arguments, but in any case the complexity of the lattice construction is reduced in this way from order $N^2$ to order $N$.

To test our random lattice construction we have measured the average link length $\langle \ell \rangle$, and the (normalized) distribution of coordination numbers $P(q)$, which can be compared with the exact results given in Refs.[2, 25]. Our results for three different realizations with $N = 10\,000$ sites and one with $N = 80\,000$ sites are collected in Table 2. Notice that for $N = 80\,000$ a single site with coordination number $q = 14$ would give $P(14) = 0.0000125$. Compared with the exact number we thus expect this to happen on the average only every 4th realization. Similarly, a site with $q = 15$ should occur only every 40th realization of a $N = 80\,000$ lattice. The average coordination number $\bar{q}$ was always exactly six, as it should be for periodic boundary conditions (i.e., on a torus) by Euler's theorem, $N - N_L + N_T = \chi$. Here $N_L$ and $N_T$ are the number of links and triangles satisfying $2N_L = 3N_T$, and $\chi$ is the Euler characteristic ($= 0$ for a torus and $= 2$ for a sphere). This implies $\bar{q} = 2N_L/N = 6(1 - \chi/N)$ and thus $\bar{q} = 6$ for the topology of a torus, while for the topology of a sphere there is a correction term $(1 - 2/N)$.

Table 2: *Coordination number distribution $P(q)$ and average link length $\langle \ell \rangle$ for the three realizations of $N = 10\,000$ and for $N = 80\,000$. The exact numbers for $P(q)$ are taken from Ref. [25], and for $\langle \ell \rangle = 32/9\pi$ see Ref. [2].*

| $q$ | $P(q)$ | | | | |
|---|---|---|---|---|---|
| | $N = 10\,000$ | $N = 10\,000$ | $N = 10\,000$ | $N = 80\,000$ | exact |
| 3 | 0.0107 | 0.0112 | 0.0127 | 0.0107375 | 0.0127(8) |
| 4 | 0.1033 | 0.1077 | 0.1090 | 0.1079625 | 0.1077(11) |
| 5 | 0.2535 | 0.2519 | 0.2573 | 0.2575125 | 0.258(2) |
| 6 | 0.3052 | 0.3037 | 0.2923 | 0.2952500 | 0.294(3) |
| 7 | 0.2048 | 0.1991 | 0.1982 | 0.2000000 | 0.198(3) |
| 8 | 0.0889 | 0.0903 | 0.0889 | 0.0904375 | 0.090(20) |
| 9 | 0.0258 | 0.0267 | 0.0313 | 0.0297500 | 0.0288(7) |
| 10 | 0.0068 | 0.0064 | 0.0084 | 0.0065500 | 0.00695(20) |
| 11 | 0.0010 | 0.0026 | 0.0016 | 0.0015000 | 0.00153(9) |
| 12 | 0.0000 | 0.0003 | 0.0002 | 0.0002750 | 0.00024(3) |
| 13 | 0.0000 | 0.0001 | 0.0001 | 0.0000250 | 0.000029(5) |
| 14 | 0.0000 | 0.0000 | 0.0000 | 0.0000000 | 0.0000028(4) |
| $\langle \ell \rangle$ | 1.1297(32) | 1.1296(33) | 1.1310(33) | 1.1312(12) | 1.131768... |

**B. Model:** We use the standard partition function of the Ising model,

$$Z = \sum_{\{s_i\}} e^{-KE}; \quad E = -\sum_{\langle ij \rangle} s_i s_j; \quad s_i = \pm 1, \qquad (1)$$

where $K = J/k_B T > 0$ is the inverse temperature in natural units, and $\langle ij \rangle$ denotes nearest-neighbor links of our two-dimensional random lattices with periodic boundary conditions. In (1) we have adopted the convention of Espriu *at al.* [12] and assigned each link the same weight.

**C. Simulation:** It is by now well known that the problem of critical slowing down of local update algorithms at continuous phase transitions can be overcome by non-local update algorithms in which whole clusters of spins are flipped in a coherent way [20]. It is intuitively clear that this leads to a much more efficient sampling of long wavelength fluctuations than in local update schemes. Currently there are two related formulations available. First, the Swendsen-Wang (SW) formulation [18], in which the whole lattice is decomposed into clusters, and second, Wolff's single cluster (1C) formulation [19], which is based on the generation of a single cluster in each step. Various tests of these cluster algorithms, in particular for the Ising model on two- and three-dimensional regular lattices, have clearly demonstrated that critical slowing down is significantly reduced [26–31]. Both formulations are easily implemented on random lattices as well. The only difference is that the coordination number now varies from site to site. In our simulations we have chosen the 1C formulation. This choice was motivated by comparative studies of the two cluster algorithms on regular cubic lattices which favor the 1C formulation. This is very pronounced in three dimensions [26, 31], but also in two dimensions we expect autocorrelation times that are about $2 - 3$ times smaller than for the SW formulation [26, 27].

For the first set of simulations at criticality we generated random lattices with $N = 5\,000, 10\,000, 20\,000, 40\,000,$ and $80\,000$ sites. For later use we adopt the notation for regular lattices and define a linear lattice size $L$ by $L = \sqrt{N}$. To investigate the dependence of thermal averages on different realizations for fixed $N$, we considered three randomly chosen realizations for $N = 5\,000$ and $10\,000$, and two for $N = 20\,000$, respectively. All runs were performed at $K_0 = 0.263$, the estimate of $K_c$ as obtained by Espriu *et al.* [12]. From $50\,000$ to $150\,000$ clusters were discarded to reach equilibrium from an initially completely disordered state, and a further $4 \times 10^6$ clusters were generated for measurements. Every $10th$ cluster the energy per spin, $e = E/N$, and the magnetization per spin, $m = \sum_i s_i/N$, were measured and recorded in a time series file. The mean cluster size $\langle |C| \rangle$ was obtained from all clusters.

Our second set of data consists of simulations in the disordered phase. Here we used one random lattice with $N = 40\,000$ sites to generate time series for $e$, $m$, and $|C|$ at $K = 0.22, 0.225, 0.23, 0.235, 0.24, 0.245, 0.25$ and $0.26$, and another lattice with $N = 80\,000$ sites at $K = 0.252$ and $0.254$. The statistics is the same as in the FSS region.

**D. Update dynamics:** To estimate the autocorrelation time of the measurements of $e$ and $\chi = KNm^2$ we used two methods. First, by measuring the (normalized) autocorrelation function $A(k) = \langle O(k); O(0)\rangle/\langle O(0); O(0)\rangle$ with $\langle O(k); O(0)\rangle \equiv \langle O(k)O(0)\rangle - \langle O(k)\rangle\langle O(0)\rangle$, we computed directly the integrated autocorrelation time $\hat{\tau} = 1/2 + \sum_{k=1} A(k)$, using a self-consistent upper cutoff [32] of $k_{\max} = 6\hat{\tau}$. For a rough error estimate we used the a priori formula [32] $\epsilon_{\hat{\tau}} = \sqrt{2(2k_{\max}+1)/N_m}\hat{\tau}$, where $N_m$ is the number of measurements. Second, we used the fact that $\hat{\tau}$ enters the error estimate $\epsilon^2 = \sigma^2 2\hat{\tau}/N_m$ for the mean $\overline{O}$ of $N_m$ correlated measurements of variance $\sigma^2 = \langle O; O\rangle$, and determined $\epsilon^2$ by blocking procedures. Using 8000 blocks of 50 measurements each we obtained agreement with the direct method at a 1-2% level. For a smaller number of blocks (3200 or 800), we observed a clear increase of the fluctuations around the directly obtained values of $\hat{\tau}$ consistent with an inverse square root behavior. Our results are compiled in Table 3. We see that the integrated autocorrelation times of the measurements of $e$ and $\chi$ are of the order of $\hat{\tau}_e \approx 0.8 - 1.3$ and $\hat{\tau}_\chi \approx 0.7 - 0.9$. Since completely uncorrelated data would give $\hat{\tau} = 0.5$, our sample thus effectively consists of about 200 000 uncorrelated measurements.

Table 3: *Integrated autocorrelation times of energy and susceptibility at the simulation point $K_0 = 0.263$ ($\approx K_c$). The $\hat{\tau}^{\mathrm{Bl}}$ follow from an error analysis using 8000 blocks of 50 measurements each, while $\hat{\tau}$ is estimated directly from the autocorrelation function.*

| $N$ | $\hat{\tau}_e^{\mathrm{Bl}}$ | $\hat{\tau}_e$ | $\hat{\tau}_\chi^{\mathrm{Bl}}$ | $\hat{\tau}_\chi$ | $\langle|C|\rangle$ | $f$ | $\tau_e$ | $\tau_\chi$ |
|---|---|---|---|---|---|---|---|---|
| 5 000 | 0.79 | 0.79(1) | 0.67 | 0.68(1) | 1564.3(1.2) | 3.13 | 2.04(2) | 1.54(2) |
| 5 000 | 0.76 | 0.76(1) | 0.67 | 0.67(1) | 1642.7(1.1) | 3.29 | 2.02(2) | 1.57(2) |
| 5 000 | 0.78 | 0.78(1) | 0.70 | 0.68(1) | 1621.0(1.1) | 3.24 | 2.05(2) | 1.58(2) |
| 10 000 | 0.93 | 0.94(1) | 0.74 | 0.75(1) | 2761.2(2.3) | 2.76 | 2.27(2) | 1.54(2) |
| 10 000 | 0.85 | 0.86(1) | 0.71 | 0.71(1) | 2992.5(2.3) | 2.99 | 2.14(2) | 1.57(2) |
| 10 000 | 0.80 | 0.80(1) | 0.70 | 0.69(1) | 3163.6(2.2) | 3.16 | 2.12(2) | 1.47(2) |
| 20 000 | 0.92 | 0.92(1) | 0.78 | 0.77(1) | 5743.7(4.8) | 2.87 | 2.31(2) | 1.68(2) |
| 20 000 | 0.92 | 0.94(1) | 0.75 | 0.76(1) | 5666.3(4.2) | 2.83 | 2.34(2) | 1.69(2) |
| 40 000 | 1.04 | 1.06(1) | 0.82 | 0.82(1) | 10693.7(8.1) | 2.67 | 2.54(3) | 1.64(2) |
| 80 000 | 1.26 | 1.28(2) | 0.93 | 0.92(1) | 18810.0(18.) | 2.35 | 2.80(3) | 1.78(2) |

While this properly characterizes the statistics of our simulations, the numbers for $\hat{\tau}$ of a single-cluster simulation are not yet well suited for a comparison with other update algorithms or even single-cluster simulations on regular lattices. To get a comparative work-estimate, the usual procedure [19] is to convert the $\hat{\tau}$ by multiplying with a factor $f = 10\langle|C|\rangle/N$ ($\approx 3$ in our case) to a scale where measurements are taken after every spin has been flipped once (similar to, e.g., Metropolis simulations). In our case, however, the measured $\hat{\tau}$'s are too small to justify this simple rescaling procedure, since non-linearities caused by the discreteness of MC time would lead to quite severe overestimates. This follows by observing that $A(k)$ is in general a convex function and that $\hat{\tau}$ can be interpreted as the trapezoidal approximation of the area under this curve. Increasing the interval between measurements by a factor of 5 or 10, say, corresponds to measuring only every 5*th* or 10*th* point of the curve one would get by taking measurements every iteration. If at the scale of the less frequent measurements $\hat{\tau} \approx 1$, then the trapezoidal approximation becomes obviously poor and overestimates the true area. To see this more explicitly we adopt the usual assumption that the autocorrelation function can be written as a sum of exponentials, $A(k) = \sum_n a_n \exp(-k/\hat{\tau}_n)$, with exponential autocorrelation times $\hat{\tau}_n$ and amplitudes $a_n$ satisfying $\sum_n a_n = A(0) = 1$. Each exponential contributes to $\hat{\tau}$ a term $\frac{1}{2}a_n \coth(1/2\hat{\tau}_n) = a_n\hat{\tau}_n[1 + 1/12\hat{\tau}_n^2 + \ldots]$. While the $\hat{\tau}_n$ do get simply rescaled when changing the measurement interval, the nonlinear relationship with $\hat{\tau}$ clearly shows that such a simple procedure cannot work for integrated autocorrelation times of the order unity.

To circumvent this problem we used the following method. We first performed fits to the Ansatz $A(k) = a\exp(-k/\hat{\tau}_0) + (1-a)\exp(-k/\hat{\tau}_1)$, and then used this function to sum the integrated autocorrelation time at interpolated $k$ values of spacing $\Delta k = 0.1$. Finally we converted these numbers to the usual "Metropolis" scale. For one realization each of the $N = 5\,000, 10\,000$, and 20 000 lattices we have repeated the runs with measurements taken after every cluster flip. This amounts to performing ten times more measurements and thus computing the interpolated values of $A(k)$ directly. On the basis of these tests we are quite sure that the interpolation method gives only small overestimates of the order $1 - 2\%$ for the energy and about 5% for the susceptibility, respectively.

To conclude this discussion, when aiming at an accurate determination of autocorrelation times with small systematic errors, it is advantageous to perform many measurements per $\hat{\tau}$. The accuracy of static quantities, however, is hardly improved by more frequent, but strongly correlated measurements.

In fact, taking into account the time spent for the measurements, it is usually even more efficient to adjust the interval between measurements such that $\hat{\tau} \approx 1$.

The numbers in Table 3 obtained in this way are very similar to results for the regular square ($sq$) lattice [26, 27]. If we fit the data for the three largest lattices to a power law, $\tau \propto L^z$, we obtain $\tau_e = 0.62(6) L^{0.27(2)}$, and $\tau_\chi = 1.2(2) L^{0.07(2)}$, respectively. Of course, we have not enough data points and our lattices are too small to exclude other scaling forms. In particular, we get also very good fits to a logarithmic scaling, $\tau = a + b \ln L$, as claimed for the $sq$ lattice [26, 27, 33], but even fits to a linear scaling, $\tau = a + bL$, are satisfactory. Explicitly we obtain from the logarithmic fits $\tau_e = -1.0(3) + 0.68(5) \ln L$ and $\tau_\chi = 1.1(2) + 0.11(4) \ln L$, and from the linear fits $\tau_e = 1.84(5) + 0.0034(3) L$ and $\tau_\chi = 1.58(4) + 0.0006(2) L$. For an illustration see Fig. 1.

**E. Observables:** From the time series of $e$ and $m$ at $K_0 = 0.263$ it is straightforward to compute in the FSS region various quantities at nearby values of $K_0$ by standard reweighting [21]. To estimate the statistical errors, the time-series data was split into 20 bins, which were jack-knived [34] to decrease the bias in the analysis of reweighted data.

In this way we determined the temperature dependence of the magnetic fourth-order Binder parameter [35],

$$U_L(K) = 1 - \frac{\langle m^4 \rangle}{3\langle m^2 \rangle^2}, \tag{2}$$

whose curves for different lattice sizes $L = \sqrt{N}$ should cross for large $L$ in the unique point $(K_c, U^*)$ (up to correction terms). We analyzed the maxima, $\chi'_{\max}$, of the (finite lattice) susceptibility,

$$\chi'(K) = K N (\langle m^2 \rangle - \langle |m| \rangle^2), \tag{3}$$

and studied the (finite lattice) magnetizations at their points of inflection, $\langle |m| \rangle|_{\inf}$. These points follow from the maxima of $d\langle |m| \rangle/dK$, which can be conveniently computed by using the fluctuation formula

$$\frac{d\langle |m| \rangle}{dK} = \langle |m| \rangle \langle E \rangle - \langle |m| E \rangle. \tag{4}$$

Useful scaling information can also be extracted from the logarithmic derivatives [36]

$$\frac{d \ln \langle |m| \rangle}{dK} = \langle E \rangle - \frac{\langle |m| E \rangle}{\langle |m| \rangle}, \tag{5}$$

and

$$\frac{d \ln \langle m^2 \rangle}{dK} = \langle E \rangle - \frac{\langle m^2 E \rangle}{\langle m^2 \rangle}. \tag{6}$$

We further looked at the maxima of the specific heat,

$$C(K) = K^2 N (\langle e^2 \rangle - \langle e \rangle^2), \tag{7}$$

and at the minima of the energetic fourth-order parameter

$$V_L(K) = 1 - \frac{\langle e^4 \rangle}{3 \langle e^2 \rangle^2}. \tag{8}$$

Note that this ratio is usually considered only at first-order phase transitions [37]. As will be demonstrated below, however, it carries useful information at a continuous phase transition as well.

In the simulations in the disordered phase we concentrated on the approach to criticality of the specific heat and the susceptibility, as defined in (3) or, since $\langle m \rangle = 0$, more properly as

$$\chi(K) = K N \langle m^2 \rangle. \tag{9}$$

For the latter definition an improved estimator is available [26], being simply the average cluster size,

$$\chi_{\text{imp}}(K) = K \langle |C| \rangle. \tag{10}$$

## III. Results in the finite-size scaling region

In this section we describe the analysis of our data near criticality, using reweighting techniques [21] and FSS ideas [22].

**A. Binder parameter and estimates for $K_c$, $U^*$ and $\nu$:** It is well known [35] that the $U_L(K)$ curves for different lattice sizes $L$ should intersect around $(K_c, U^*)$ with slopes $U'_L \equiv dU_L/dK \propto L^{1/\nu}$, where $U^*$ is the (universal) "renormalized charge" and $\nu$ is the critical exponent of the correlation length. More precisely, due to corrections to the leading FSS behavior, the curves for $L$ and $L'$ should cross in points $K^\times = K^\times(L, L')$, which approach $K_c$ for large lattice sizes. Our results for $K^\times(L, L')$ obtained by reweighting the primary data at $K_0 = 0.263$ are collected in Table 4. By looking through the table we observe that there are rather strong fluctuations between different replicas for a fixed number of sites $N$. Even though these fluctuations decrease with increasing $N$, with the present data it is impossible to apply the standard extrapolation formula [35] for estimating $K_c$. Taking hence as our final estimate for $K_c$ the average of the five values for $K^\times(L, L')$ from the three largest lattices, we obtain

$$K_c = 0.2630 \pm 0.0002. \quad (11)$$

This value is in good agreement with the estimates in Ref.[12] from high-temperature series expansions ($K_c \approx 0.26303$) and MC simulations in the disordered phase ($K_c = 0.2631(3)$). As already noted in Ref.[12] the value for $K_c$ is very close to the exact critical coupling $\ln(3)/4 = 0.27465\ldots$ of the regular triangular lattice whose coordination number is also $q = 6$. The (rough) error estimate in (11) should also reflect the fluctuations caused by the different random lattice realizations for fixed $N$, which in our case are much bigger than the statistical errors.

For the same reason, the estimates of $U^*$ in Table 4 show stronger fluctuations than our statistics would suggest. Taking the average over all lattice sizes and replicas at our estimate of the critical point, $K_c = 0.2630$, we obtain

$$U^* = 0.6123 \pm 0.0025. \quad (12)$$

The corresponding values at $K = K_c - 0.0002$ and $K = K_c + 0.0002$ are $U^* = 0.6054(25)$ and $U^* = 0.6183(28)$, so that taking into account the uncertainties in $K_c$ the error bar in (12) should probably be increased to 0.0070. This then would be consistent with an average over the five lower entries in Table 4, which gives $U^* = 0.6176(60)$. Our value for $U^*$ is practically indistinguishable from MC estimates for the regular $sq$ lattice which are $U^* = 0.615(10)$ [38] and $U^* = 0.611(1)$ [33]. This good agreement may be taken as a first indication of lattice universality.

Table 4: *Estimates of $\nu_{\mathrm{eff}}(L, L')$, $K^\times(L, L')$, and $U^* \approx U_L(K^\times(L, L'))$ from the magnetic fourth-order Binder parameter for pairs of lattices of size $L$ and $L'$.*

| $L \setminus L'$ | 10000 | 10000 | 10000 | 20000 | 20000 | 40000 | 80000 |
|---|---|---|---|---|---|---|---|
| 5000 | 1.008(16) 0.263744(43) 0.61747(88) | 0.993(17) 0.262083(40) 0.5814(14) | 1.004(48) 0.260813(81) 0.5417(34) | 1.002(13) 0.262172(17) 0.58378(74) | 0.980(14) 0.262346(24) 0.58827(86) | 1.010(12) 0.262459(16) 0.59107(66) | 1.0017(45) 0.628107(67) 0.59931(40) |
| 5000 | 1.012(20) 0.264840(29) 0.63764(33) | 1.023(20) 0.263155(23) 0.61498(46) | 1.039(32) 0.261847(46) 0.5865(14) | 1.020(11) 0.262615(13) 0.60457(38) | 1.004(12) 0.262780(16) 0.60794(41) | 1.017(10) 0.262704(10) 0.60641(32) | 1.0083(49) 0.2629592(54) 0.61141(21) |
| 5000 | 1.012(14) 0.264493(30) 0.63238(41) | 1.020(13) 0.262811(31) 0.60576(75) | 1.042(34) 0.261506(52) 0.5733(18) | 1.017(11) 0.262474(16) 0.59849(53) | 0.999(12) 0.262641(15) 0.60220(44) | 1.0163(97) 0.262626(11) 0.60187(38) | 1.0068(41) 0.2629123(49) 0.60782(23) |
| 10000 | | | | 0.979(95) 0.26104(10) 0.5061(64) | 0.918(58) 0.261414(66) 0.5275(39) | 1.047(31) 0.262064(27) 0.5600(14) | 1.0039(69) 0.2626603(75) 0.58475(55) |
| 10000 | | | | 1.011(27) 0.262237(33) 0.5871(14) | 0.974(22) 0.262527(36) 0.5971(13) | 1.015(15) 0.262572(13) 0.59853(57) | 1.0028(46) 0.2629284(66) 0.60907(34) |
| 10000 | | | | 1.017(15) 0.263157(27) 0.62359(70) | 0.997(17) 0.263420(27) 0.62876(62) | 1.0111(98) 0.262954(12) 0.61914(42) | 1.0045(56) 0.2631351(70) 0.62314(28) |
| 20000 | | | | | | 1.010(20) 0.262811(20) 0.61220(83) | 0.9984(62) 0.2631289(75) 0.62277(31) |
| 20000 | | | | | | 1.055(39) 0.262607(29) 0.6007(15) | 1.0106(90) 0.2630502(77) 0.61781(39) |
| 40000 | | | | | | | 1.002(16) 0.263352(15) 0.63431(54) |

To extract the critical exponent $\nu$ several methods are possible. One could, e.g., analyze the scaling of the slopes $U'_L(K) = dU_L/dK \propto L^{1/\nu}$ at $K = K_c$, or more generally at any sequence of $K$ values for which the scaling variable $x = (K - K_c)L^{1/\nu}$ is constant. Examples are the locations of the specific heat or susceptibility maxima which are expected to scale as $K_{\max}(L) = K_c + aL^{-1/\nu}$, with $a$ being a constant. Another convenient choice of such a sequence are the points where $U'_L(K)$ is maximal. In this case one gets the desired slopes $U'_{\max}(L)$ directly, without explicit knowledge of the corresponding $K$ values.

We have tried all of these possibilities, but not all gave sensitive results in our case. The errors on $U'_L$ at fixed $K = K_c$ are clearly dominated by the replica fluctuations. Without further simulations to increase the replica statistics it is then impossible to obtain reliable fits. The second, self-consistent method turned out to be much more suited for our problem. With the simplest choice of the maxima of $U'_L$, however, we run into the problem that they lie too far away from our simulation point, thus allowing no safe reweighting. Choosing as sequence of $K$-values the locations of the specific-heat or susceptibility maxima, $K^C_{\max}(L)$ or $K^{\chi'}_{\max}(L)$ (see Table 5 below), or the inflection points of the magnetization, $K^{(|m|)}_{\inf}(L)$ (see Table 6 below), we obtain the fits shown in Fig. 2. From the inverse of the slopes we read off

$$\nu = 1.021 \pm 0.033 \qquad (\text{at } K^C_{\max}), \tag{13}$$

$$\nu = 0.996 \pm 0.028 \qquad (\text{at } K^{\chi'}_{\max}), \tag{14}$$

$$\nu = 1.020 \pm 0.020 \qquad (\text{at } K^{(|m|)}_{\inf}), \tag{15}$$

in very good agreement with the Onsager value for regular lattices, $\nu = 1$. The quality of the fits, however, is relatively poor.

We therefore used finally yet another approach which is based on the effective exponents

$$\nu_{\text{eff}} = \frac{\ln(L'/L)}{\ln\left(U'_{L'}(K^\times)/U'_L(K^\times)\right)}. \tag{16}$$

Our results for $\nu_{\text{eff}}$ are again collected in Table 4 and plotted as open circles in Fig. 3, where the $x$-axis corresponds to the 38 possible combinations of $L$ and $L'$. We see that within the error bars all entries are compatible with $\nu = 1$. The average over all entries gives

$$\nu = 1.008 \pm 0.022 \qquad (\text{effective } \nu\text{'s}), \tag{17}$$

where the error estimate is the standard deviation of the $\nu_{\text{eff}}$. If we take only the 9 crossing points of the $N = 80\,000$ lattice with all other lattices into account (filled circles in Fig. 3), then we obtain

$$\nu = 1.0043 \pm 0.0036 \qquad (\text{effective } \nu\text{'s}). \tag{18}$$

We can thus conclude that our estimates of the exponent $\nu$ for random lattices are fully consistent (at a 0.5% level) with the exact regular lattice value of $\nu = 1$.

Assuming $\nu = 1$, we can use the asymptotic FSS behavior of the pseudo-transition points, e.g., $K^C_{\max}(L) = K_c + aL^{-1/\nu}$, to obtain further estimates of the critical coupling $K_c$ from linear fits in $1/L$. Our results from fits to the data (see Tables 5 and 6 below) of the three largest lattices are

$$K_c = 0.26295 \pm 0.00033 \qquad (\text{from } K^C_{\max}), \tag{19}$$

$$K_c = 0.262947 \pm 0.000077 \qquad (\text{from } K^{\chi'}_{\max}), \tag{20}$$

$$K_c = 0.26304 \pm 0.00014 \qquad (\text{from } K^{(|m|)}_{\inf}), \tag{21}$$

in good agreement with the estimate (11) based on the Binder parameter intersections.

**B. Susceptibility and $\gamma/\nu$:** To extract the ratio of exponents $\gamma/\nu$ we used that the maxima of the susceptibility should scale for sufficiently large $L$ like

$$\chi'_{\max}(L) = \chi'(K^{\chi'}_{\max}(L), L) = AL^{\gamma/\nu}. \tag{22}$$

Our results for $K^{\chi'}_{\max}$ and $\chi'_{\max}$ are collected in Table 5, and the $\chi'_{\max}$ are plotted vs $L$ on a log-log scale in Fig. 4. The straight line fit shown in Fig. 4 gives

$$\gamma/\nu = 1.7503 \pm 0.0059, \tag{23}$$

with an amplitude $A = 0.02491(67)$, and a goodness-of-fit parameter [39] $Q = 0.035$. This is again in excellent agreement with the exact value for the two-dimensional Ising model on a regular lattice, $\gamma/\nu = 1.75$. Even though the $Q$ value of the fit is quite low, we do not see in Fig. 4 any trend with increasing lattice size. In fact, if we discard the data for $N = 5000$, the fit yields $\gamma/\nu = 1.7468(91)$ with $Q = 0.015$, and if we further discard the data for $N = 10000$, we obtain $\gamma/\nu = 1.735(19)$ with $Q = 0.005$. Constrained fits with $\gamma/\nu = 1.75$ held fixed at its theoretical value, are equally acceptable and yield for the amplitude $A = 0.024938(45)$ with $Q = 0.071$, or, discarding the $N = 5000$ data, $A = 0.024952(59)$ with $Q = 0.037$. We can thus conclude that universality also holds as far as $\gamma/\nu$ is concerned.

Notice that in these fits (whether linear or not) it does not matter whether we fit over all 10 data points or first compute the weighted replica averages for $N = 5000, 10000$ and $20000$ and then fit over 5 data points. It is easy to show [40] that the results must be identically the same, apart from the $Q$ values. Here and in the following we always quote the $Q$ value for first computing the replica averages.

Table 5: *Extrema for the (finite lattice) susceptibility ($\chi'_{\max}$), the specific heat ($C_{\max}$), and the the energy moment ratio ($V_{\min}$), together with their respective pseudo critical couplings.*

| $N$ | $K^{\chi'}_{\max}$ | $\chi'_{\max}$ | $K^{C}_{\max}$ | $C_{\max}$ | $K^{V}_{\min}$ | $V_{\min}$ |
|---|---|---|---|---|---|---|
| 5000 | 0.259772(39) | 42.77(19) | 0.262041(66) | 2.0109(53) | 0.26043(17) | 0.664378(12) |
| 5000 | 0.259305(63) | 43.09(29) | 0.26141(12) | 2.017(10) | 0.25940(59) | 0.664321(33) |
| 5000 | 0.259448(39) | 43.09(18) | 0.26164(11) | 2.0138(71) | 0.26005(22) | 0.664354(22) |
| 10000 | 0.260930(28) | 79.27(29) | 0.262421(49) | 2.1600(73) | 0.261532(93) | 0.6654595(53) |
| 10000 | 0.260446(30) | 78.61(34) | 0.262092(78) | 2.1374(68) | 0.26131(11) | 0.6654720(74) |
| 10000 | 0.260037(48) | 78.17(65) | 0.261723(73) | 2.1345(89) | 0.260954(88) | 0.6654670(81) |
| 20000 | 0.260871(55) | 145.2(1.4) | 0.262086(86) | 2.276(13) | 0.26160(16) | 0.6660347(66) |
| 20000 | 0.261023(30) | 146.3(1.2) | 0.262198(11) | 2.311(18) | 0.26150(14) | 0.6660199(82) |
| 40000 | 0.261526(33) | 259.1(2.5) | 0.262310(84) | 2.421(22) | 0.262081(90) | 0.6663340(43) |
| 80000 | 0.261991(41) | 492.2(6.1) | 0.2623(28) | 2.600(97) | 0.26149(83) | 0.666481(11) |

**C. Magnetization and $\beta/\nu$:** The standard way to extract the exponent ratio $\beta/\nu$ is to consider the FSS of the magnetization at $K_c$,

$$\langle |m| \rangle (K_c) \propto L^{-\beta/\nu}. \qquad (24)$$

We tried this also here, but due to the replica fluctuations the resulting scaling curve was difficult to analyze. As a solution to this problem we decided to study the scaling behavior of $\langle |m| \rangle$ at the point of inflection, i.e., at the point where $d\langle |m| \rangle / dK$ is maximal. Since these points should scale as usual, $(K^{\langle |m| \rangle}_{\inf} - K_c) L^{1/\nu} \equiv t L^{1/\nu} = const$, we expect

$$\langle |m| \rangle |_{\inf} = L^{-\beta/\nu} f(t L^{1/\nu}) \propto L^{-\beta/\nu}, \qquad (25)$$

and, since a derivative with respect to $K$ picks up a factor $L^{1/\nu}$ from the argument of the scaling function $f$,

$$\frac{d\langle |m| \rangle}{dK} |_{\max} = L^{-\beta/\nu + 1/\nu} f'(t L^{1/\nu}) \propto L^{(1-\beta)/\nu}. \qquad (26)$$

Consequently, the scaling of the maxima of the logarithmic derivatives (5) and (6) should be given by

$$\frac{d \ln \langle |m| \rangle}{dK} |_{\max} = \frac{d\langle |m| \rangle / dK}{\langle |m| \rangle} |_{\max} \propto L^{1/\nu}, \qquad (27)$$

Table 6: *Inflection points $K^{\langle |m| \rangle}_{\inf}$ of the magnetization, and $\langle |m| \rangle$ and $d\langle |m| \rangle / dK$ at $K^{\langle |m| \rangle}_{\inf}$. Also given are the extrema of the logarithmic derivatives (5) and (6), and their respective pseudo critical couplings.*

| $N$ | $K^{\langle |m| \rangle}_{\inf}$ | $\langle |m| \rangle |_{\inf}$ | $\frac{d\langle |m| \rangle}{dK}|_{\max}$ | $K^{\ln \langle |m| \rangle}_{\max}$ | $\frac{d \ln \langle |m| \rangle}{dK}|_{\max}$ | $K^{\ln \langle m^2 \rangle}_{\max}$ | $\frac{d \ln \langle m^2 \rangle}{dK}|_{\max}$ |
|---|---|---|---|---|---|---|---|
| 5000 | 0.260943(47) | 0.4474(21) | 45.73(19) | 0.25859(26) | 114.9(1.8) | 0.25811(27) | 192.8(2.9) |
| 5000 | 0.26034(10) | 0.4377(51) | 46.65(44) | 0.2571(12) | 124.1(6.7) | 0.2564(12) | 211(13) |
| 5000 | 0.260525(71) | 0.4409(35) | 46.43(31) | 0.25800(42) | 119.5(3.2) | 0.25755(42) | 200.1(5.4) |
| 10000 | 0.261709(38) | 0.4247(23) | 62.46(27) | 0.26014(14) | 164.9(2.5) | 0.25979(19) | 277.2(4.6) |
| 10000 | 0.261298(52) | 0.4289(33) | 61.76(35) | 0.25987(16) | 159.5(3.2) | 0.25955(21) | 267.0(5.8) |
| 10000 | 0.260942(44) | 0.4293(28) | 61.61(51) | 0.25929(29) | 160.4(3.3) | 0.25909(24) | 266.5(4.8) |
| 20000 | 0.261499(93) | 0.4039(84) | 83.7(1.1) | 0.25832(67) | 288(57) | 0.25782(72) | 485(96) |
| 20000 | 0.261529(62) | 0.3985(58) | 85.7(1.2) | 0.26038(14) | 244.2(8.4) | 0.26014(16) | 409(14) |
| 40000 | 0.261979(33) | 0.3980(45) | 111.5(1.7) | 0.261359(86) | 305.8(9.2) | 0.261247(83) | 512(15) |
| 80000 | 0.262244(92) | 0.359(15) | 156.2(2.8) | 0.26098(83) | 576(137) | 0.2606(12) | 972(276) |

and

$$\frac{d \ln \langle m^2 \rangle}{dK} |_{\max} = \frac{d\langle m^2 \rangle / dK}{\langle m^2 \rangle} |_{\max} \propto L^{1/\nu}, \qquad (28)$$

thus providing another means to estimate the correlation length exponent $\nu$.

Our data for these quantities is given in Table 6. The scaling of $\langle |m| \rangle$ at the inflection point is shown in the log-log plot of Fig. 5. The linear fit through all data points gives

$$\beta/\nu = 0.1208 \pm 0.0092, \qquad (29)$$

with $Q = 0.10$. This value is again perfectly compatible with the exact result for regular lattices, $\beta/\nu = 0.125$. Omitting the data of the smallest lattice, we obtain $\beta/\nu = 0.122(15)$, with $Q = 0.04$.

The scaling of the maxima of $d\langle |m| \rangle / dK$ is shown in Fig. 6. Here we obtain from the fit

$$(1 - \beta)/\nu = 0.8704 \pm 0.0081 \qquad (30)$$

with $Q = 0.39$, in agreement with the regular lattice result $(1 - \beta)/\nu = 0.875$. Omitting the data of the smallest lattice our estimate even improves to $(1 - \beta)/\nu = 0.875(13)$ with $Q = 0.24$.

Finally, using the maxima of the logarithmic derivatives (5) and (6) we obtain two further estimates for $\nu$. From the linear fits in the log-log plots shown in Fig. 7 we read off

$$\nu = 1.037 \pm 0.031 \qquad (31)$$

with $Q = 0.06$, and
$$\nu = 1.042 \pm 0.030 \qquad (32)$$
with $Q = 0.04$, respectively. The two values for $\nu$ are fully consistent with our previous estimates and with the regular lattice result of $\nu = 1$.

**D. Specific heat and $\alpha/\nu$:** We now turn to the specific heat which is usually the most difficult quantity to analyze. The reason is that, compared to the susceptibility, the critical divergence is much weaker and regular background terms become important. Recalling our result $\nu \approx 1$ and assuming hyperscaling to be valid for the random lattice as well, we expect $\alpha = 2 - d\nu \approx 0$, as for a regular lattice. The corresponding FSS prediction is then
$$C_{\max}(L) = C(K^C_{\max}(L), L) = B_0 + B_1 \ln L. \qquad (33)$$
The semi-log plot in Fig. 8 clearly demonstrates that our data in Table 5 is consistent with this prediction. A linear fit through all data points gives $B_0 = 0.346(52)$ and $B_1 = 0.391(12)$ with $Q = 0.84$. It should be remarked, however, that the confirmation of $\alpha = 0(\log)$ is not really conclusive. Due to the small range over which $C_{\max}$ varies we can fit the data also with a simple power-law Ansatz $C_{\max} \propto L^{\alpha/\nu}$, yielding $\alpha/\nu = 0.1824(53)$ with $Q = 0.93$; see Fig. 8. Discarding first only the data points for $N = 5000$ and then also those for $N = 10000$, we obtain $\alpha/\nu = 0.180(10)$ with $Q = 0.83$ and $\alpha/\nu = 0.168(27)$ with $Q = 0.72$, respectively. There is thus a small downward trend, but our data is obviously also consistent with a power-law Ansatz. We also tried a non-linear three-parameter fit to the more reasonable Ansatz $C_{\max} = b_0 + b_1 L^{\alpha/\nu}$. As a result we then obtain an exponent ratio $\alpha/\nu = 0.17(16)$ consistent with zero, but the errors on all three parameters are much too large to draw a firm conclusion from such a fit.

To convince ourselves that these problems are not a special property of random lattices, we have compared our results with similar fits for the Ising model on a regular $sq$ lattice, employing its known analytical solution for $L \times L$ lattices with periodic boundary conditions [41]. The results for $K^C_{\max}$ and $C_{\max}$ for various $L$ are collected in Table 7. The values of $L$ that roughly correspond to our random lattice sizes are $L = 80, 100, 140, 200$ and $280$. If we fit the corresponding 5 values of $C_{\max}$ to the Ansatz (33), we obtain $B_0 = 0.1831$ and $B_1 = 0.4976$, in reasonable agreement with the exact results [41] $B_0 = 0.201359\ldots$ and $B_1 = 8K_c^2/\pi = 2(\ln(1 + \sqrt{2}))^2/\pi = 0.494358\ldots$.

The small discrepancies can be attributed to the neglected correction terms of the type $B_2 L^{-1} \ln L + B_3 L^{-1}$. While $B_2 = \frac{3}{2}\sqrt{2}B_1 \times 0.31775 = 0.333222$ is also known analytically, $B_3$ has not been worked out explicitly. Using all data in Table 7 and keeping the parameters $B_0$, $B_1$, and $B_2$ fixed at their theoretical values, we estimate numerically $B_3 \approx -2$. If we try to fit the regular lattice data for $L = 80, \ldots, 280$ to a power law $L^{\alpha/\nu}$ (which is definitely *wrong* in this case), we obtain $\alpha/\nu = 0.1941$, i.e., a value of roughly the same size as for the corresponding random lattice fit.

Table 7: *Maxima of the specific heat, $C_{\max}$, for the Ising model on a regular simple square lattice of size $L \times L$, and the corresponding pseudo critical couplings $K^C_{\max}$, as computed from the exact expressions given in Ref.[41]. The numerical errors are of the order $\pm 1$ in the last digit.*

| $L$ | $K^C_{\max}$ | $C_{\max}$ |
|---|---|---|
| 20 | 0.43323 | 1.6659 |
| 40 | 0.43685 | 2.0167 |
| 60 | 0.43811 | 2.2200 |
| 80 | 0.43874 | 2.3637 |
| 100 | 0.43912 | 2.4750 |
| 120 | 0.43938 | 2.5657 |
| 140 | 0.43957 | 2.6424 |
| 160 | 0.43971 | 2.7088 |
| 180 | 0.43981 | 2.7673 |
| 200 | 0.43990 | 2.8196 |
| 220 | 0.43997 | 2.8669 |
| 240 | 0.44003 | 2.9101 |
| 260 | 0.44008 | 2.9498 |
| 280 | 0.44012 | 2.9865 |
| 290 | 0.44014 | 3.0039 |

**E. Energy cumulant:** We also looked at the fourth-moment parameter $V_L(K)$ defined in (8), involving the energy moments. By rewriting $V_L$ as
$$V_L = 1 - \frac{1}{3} \times \frac{1 + 6\langle \delta e^2 \rangle + 4\langle \delta e^3 \rangle + \langle \delta e^4 \rangle}{1 + 2\langle \delta e^2 \rangle + \langle \delta e^2 \rangle^2}, \qquad (34)$$

with $\delta e \equiv (e - \langle e \rangle)/\langle e \rangle$, it is easy to see that in leading order

$$V_L \sim \frac{2}{3} - \frac{4}{3NK^2} \frac{C}{\langle e \rangle^2}. \qquad (35)$$

In the FSS region the energy $\langle e \rangle \approx -1.9$ varies very little. As a function of $K$ we thus expect to see a minimum in $V_L$ at roughly the point where the specific heat $C$ is maximal. As can be seen in Table 5 this is indeed the case. Assuming the usual scaling behavior of $K_{\min}^{V_L}$ and $\nu = 1$, we obtain from a linear fit $K_c = 0.26329(40)$ with $Q = 0.25$, in perfect agreement with our previous results.

Furthermore, recalling the FSS of the specific-heat maxima, we expect that the minima $V_{\min}$ of $V_L$ should scale according to $N(2/3 - V_{\min}) = b_0 + b_1 \ln L$ with $b_1 \approx 2$. In the semi-log plot of Fig. 9 we show the data for $N(2/3 - V_{\min})$ together with such a fit (with $Q = 0.29$), yielding $b_0 = 4.04(54)$ and $b_1 = 1.74(12)$. In view of the neglected corrections in (35) and the ambiguities in the fits of $C_{\max}$, the agreement with the expectation is quite satisfactory. Similarly to the specific-heat maxima, however, the data can also be fitted with an Ansatz $N(2/3 - V_{\min}) \propto L^{\alpha/\nu}$, yielding $\alpha/\nu = 0.144(10)$ and $Q = 0.43$; see Fig. 9.

## IV. Results in the disordered phase

To supplement the FSS analyzes near criticality we have performed further simulations in the disordered phase. Here we concentrated on the approach to criticality of the susceptibility and the specific heat. Most data were obtained from one random lattice with $N = 40\,000$ sites in the inverse temperature range $K = 0.22\ldots0.26$; see Table 8. The quoted autocorrelation times refer to the scale at which the measurements are taken. If this is converted to a Metropolis scale (with the unit of time set to $N$ spin flips) by multiplying with a factor $f = 10\langle |C|\rangle/N$, we obtain $\tau_e \approx 1-2$ and $\tau_\chi \approx 0.07$ (apart from the point at $K = 0.260$).

**A. Susceptibilities and exponent $\gamma$:** To analyze the data for the susceptibilities in Table 8 we have assumed the leading singular behavior as $K_c$ is approached,

$$\chi = A(K_c - K)^{-\gamma}, \qquad (36)$$

Table 8: *Results in the disordered phase for a random lattice of $N = 40\,000$ sites.*

| $K$ | $\tau_e$ | $\tau_\chi$ | $\langle e \rangle$ | $C$ | $\langle |m| \rangle$ | $\chi$ | $\chi_{\mathrm{imp}}$ | $\chi'$ |
|---|---|---|---|---|---|---|---|---|
| 0.220 | 85(7) | 12.8(4) | −1.20811(36) | 0.515(11) | 0.018850(77) | 4.901(35) | 4.9825(54) | 1.774(14) |
| 0.225 | 77(6) | 11.0(3) | −1.26321(37) | 0.589(13) | 0.02120(13) | 6.354(69) | 6.3436(65) | 2.311(25) |
| 0.230 | 67(5) | 8.0(2) | −1.32164(33) | 0.648(15) | 0.024004(88) | 8.326(52) | 8.337(11) | 3.026(19) |
| 0.235 | 60(4) | 6.1(2) | −1.38451(31) | 0.715(16) | 0.02789(15) | 11.44(11) | 11.426(14) | 4.129(39) |
| 0.240 | 50(3) | 4.06(7) | −1.45204(40) | 0.836(12) | 0.033147(93) | 16.550(84) | 16.593(29) | 6.003(33) |
| 0.245 | 35(2) | 2.61(4) | −1.52511(27) | 0.910(12) | 0.04110(12) | 26.00(15) | 26.247(34) | 9.447(59) |
| 0.250 | 24(1) | 1.49(2) | −1.60538(31) | 1.077(10) | 0.05531(11) | 47.86(17) | 48.006(87) | 17.260(63) |
| 0.260 | 3.30(5) | 0.910(7) | −1.80866(10) | 1.8356(70) | 0.21371(29) | 664.7(1.6) | 664.2(1.1) | 189.71(40) |

and performed non-linear three-parameter fits. On the one hand, this requires $K$ values that are sufficiently far away from $K_c$ to guarantee negligible finite-size effects. On the other hand, they should be sufficiently near to $K_c$ to avoid confluent and analytical correction terms in (36), which are difficult to deal with numerically [42]. Both conditions are non-universal and can only be satisfied self-consistently, relying on the goodness of the fits. Alternatively, one may rewrite the Ansatz (36) as a function of temperature $T$, or one can consider $\chi/K$ instead of $\chi$. In effect this influences the importance of the analytic correction term and thus the range of $K$- or $T$-values over which the simple Ansatz with only the leading singularity can be applied. Our results of goodness-of-fit tests for these different possibilities can be summarized as follows. If we consider $\chi/K = N\langle m^2 \rangle$, $\chi_{\mathrm{imp}}/K = \langle |C| \rangle$, or $\chi'/K$ (as in most previous works), then fits to $A(K_c - K)^{-\gamma}$ are self-consistent in an interval $K \in (0.22, 0.25)$. The data at $K = 0.26$ clearly display finite-size effects. The inclusion of the data at $K = 0.22$ seems marginal in the sense that the goodness $Q$ of the fits (or, equivalently, their chi-squared $\chi^2$) is still acceptable, but worse than for fits omitting these data. For this reason we give in Table 9 the results for both fitting ranges, corresponding to 4 and 3 degrees of freedom (DOF), respectively. We see that all estimates for $K_c$ are compatible with our FSS values of $K_c \approx 0.2630$. As expected the most precise values result from fits to the improved susceptibility, $\chi_{\mathrm{imp}}/K = \langle |C| \rangle$.

Table 9: *Results of non-linear three-parameter fits of the susceptibility data in Table 8 to the leading singularity* $A(K_c - K)^{-\gamma}$.

|  | $K = 0.220\ldots0.250$ 4 DOF | | | $K = 0.225\ldots0.250$ 3 DOF | | |
|---|---|---|---|---|---|---|
| obs. | $A$ | $K_c$ | $\gamma$ | $A$ | $K_c$ | $\gamma$ |
| $\chi_{\rm imp}/K$ | 0.0849(18) | 0.26281(10) | 1.7725(76) | 0.0836(23) | 0.26288(13) | 1.7783(99) |
|  |  | $Q = 0.90$ |  |  | $Q = 0.98$ |  |
| $\chi/K$ | 0.0793(73) | 0.26305(39) | 1.795(33) | 0.093(12) | 0.26251(48) | 1.741(44) |
|  |  | $Q = 0.44$ |  |  | $Q = 0.82$ |  |
| $\chi'/K$ | 0.0272(28) | 0.26336(43) | 1.816(37) | 0.0324(45) | 0.26276(52) | 1.757(48) |
|  |  | $Q = 0.50$ |  |  | $Q = 0.97$ |  |

This is also illustrated in a slightly different way in Fig. 10, where we plot the chi-squared per degree of freedom, $\chi^2$ per DOF, of linear least-squares fits with *fixed* $K_c$ versus $K_c$. Actually we have plotted $\sqrt{\chi^2 \text{ per DOF}}$ in order to demonstrate that the chi-squared increases quadratically as $K_c$ is varied around its optimal value.

Regarding the critical exponent $\gamma$ we observe that the estimates coming from the fits to $\chi/K = N\langle m^2 \rangle$ and $\chi'/K = N(\langle m^2 \rangle - \langle |m| \rangle^2)$ are, within error bars, fully consistent with the regular lattice value of $\gamma = 7/4 = 1.75$. The results for $\gamma$ obtained from fits to $\chi_{\rm imp}/K = \langle |C| \rangle$ overestimate this value by about $3\sigma$. Thus, taking the error estimates at face value, these results are only barely consistent with 7/4. Since in absolute terms, however, the deviation is only about 1.6%, we have not tried to improve the statistical consistency by doing a more refined error analysis (with correlation effects between parameters taken more properly into account), which usually tends to increase the error bars.

Notice finally that also $\langle |m| \rangle$ can be used to extract the exponent $\gamma$, even though this quantity should tend to zero in the infinite volume limit. The point is that, since $\chi = N\langle m^2 \rangle \sim A(K_c - K)^{-\gamma}$, we expect that $\langle |m| \rangle = \langle \sqrt{m^2} \rangle \approx \sqrt{\langle m^2 \rangle} \sim a(K_c - K)^{-\gamma/2}$, with an amplitude $a = \sqrt{A/N}$ which vanishes for large $N$. In fact, a fit in the interval $K \in [0.22, 0.25]$ yields $\gamma = 1.777(36)$ and $a = 0.0012(1)$ (with $Q = 0.41$), consistent with $\sqrt{A/N} = 0.0014(1)$, as follows by inserting $N = 40\,000$ and recalling $A = 0.079(8)$.

**B. Specific heat and exponent $\alpha$:** In view of our FSS results in Sec. III, we have assumed that $\alpha = 0$ and tried to confirm this assumption by fits of our specific heat data to a logarithmic divergence of the form

$$C = A_0 - A_1 \ln(K_c - K). \qquad (37)$$

Using all data from $K = 0.22$ to $K = 0.25$, the results of such a non-linear three-parameter fit are $A_0 = -1.07(22)$, $A_1 = 0.516(82)$, and $K_c = 0.2657(41)$, with $Q = 0.18$. Omitting the point at $K = 0.22$ we obtain $A_0 = -0.90(25)$, $A_1 = 0.455(89)$, and $K_c = 0.2631(41)$, with $Q = 0.14$. Here we have also tried to include further data from two additional runs at $K = 0.252$ and $K = 0.254$ on a lattice with 80 000 sites; see Table 10. For the susceptibility analysis we had to discard this data since, due to the very small error bars on the $\chi$'s, the susceptibility fits are much more sensitive. If the $N = 80\,000$ data was taken into account, the goodness-of-fit parameter $Q$ decreased by about two or even several orders of magnitude for fits to $\chi'$ and $\chi$ or to the more accurate $\chi_{\rm imp}$, respectively. In this data we thus saw either finite-size effects or, more likely, replica fluctuations, or a combination of both. For the specific heat, on the other hand, the relative errors are much bigger, and fits including the $N = 80\,000$ data gave almost identical results, with even improved $Q$ values ($A_0 = -1.08(15)$, $A_1 = 0.518(52)$, $K_c = 0.2657(23)$, $Q = 0.36$ for $K \in [0.220, 0.254]$, and $A_0 = -0.99(18)$, $A_1 = 0.489(60)$, $K_c = 0.2647(24)$, $Q = 0.31$ for $K \in [0.225, 0.254]$).

Table 10: *Results in the disordered phase for a random lattice of $N = 80\,000$ sites.*

| $K$ | $\hat{\tau}_e$ | $\hat{\tau}_\chi$ | $\langle e \rangle$ | $C$ | $\langle |m| \rangle$ | $\chi$ | $\chi_{\rm imp}$ | $\chi'$ |
|---|---|---|---|---|---|---|---|---|
| 0.252 | 36(2) | 2.2(2) | $-1.63893(19)$ | 1.151(16) | 0.04470(12) | 63.04(31) | 63.546(35) | 22.77(11) |
| 0.254 | 29(2) | 1.6(2) | $-1.67574(21)$ | 1.223(14) | 0.053549(91) | 90.86(29) | 91.424(55) | 32.59(11) |

The values for $K_c$ are compatible with, but considerably less accurate than our previous estimates which all gave approximately $K_c = 0.2630$. In the semi-log plot of Fig. 11(a) we therefore show $C$ vs $K_c - K$ with $K_c = 0.2630$. The solid straight line is a linear fit of the form (37) with $K_c$ held fixed at its best value. Using all data points with $K \in [0.220, 0.254]$ this yields $A_0 = -0.902(32)$ and $A_1 = 0.4544(82)$, with $Q = 0.28$.

The Fig. 11(b) shows the same data and the fit in a log-log representation. Here a pure power law would result in a straight line. Even though this is obviously not the case at $K_c = 0.2630$, fits to a pure power-law Ansatz, $C \propto (K_c - K)^{-\alpha}$, with $K_c$ as a free parameter are still acceptable ($Q = 0.26$, for $K \in [0.220, 0.254]$). The parameters, however, then take "unreasonable" values, $K_c = 0.305(18)$ and $\alpha = 1.7(5)$, and the error bars are very large. We also tried to include a constant background term by performing fits to the Ansatz $C = a_0 - a_1(K_c - K)^{-\alpha}$, with $K_c$ held fixed at values around 0.2630. This yields $\alpha$ consistent with zero, $\alpha = -0.13(9)$, albeit again with the drawback of huge error bars on all three parameters. Similar to the FSS analysis in Sec. III, also here we cannot really exclude a possible power-law scaling of the specific heat with an exponent $\alpha \neq 0$. We obtain, however, a perfectly consistent picture if we assume logarithmic scaling, that is a value of $\alpha = 0$.

## V. Concluding remarks

We have performed a fairly detailed analysis of single-cluster Monte Carlo simulations of the Ising model on two-dimensional Poissonian random lattices of Voronoi/Delaunay type. In the first set of simulations at criticality we have applied finite-size scaling methods to various quantities to extract the critical exponents of this model. At first sight our use of different quantities to estimate the same exponent might appear redundant, since the various estimates are, of course, not independent in a statistical sense. Their consistency, however, gives confidence that corrections to the asymptotic scaling behavior are very small and can safely be neglected. Direct analyses of thermodynamic measurements of the susceptibility in the disordered phase yield compatible results. From both types of simulations the results for the critical exponent $\alpha$ of the specific heat are not really conclusive, but certainly consistent with $\alpha = 0$, i.e., with a logarithmic scaling behavior. On the other hand, from our estimates for the exponents $\nu$, $\beta/\nu$, $\gamma/\nu$, and $\gamma$, we obtain strong evidence that the Ising model on two-dimensional random lattices behaves qualitatively as on regular lattices, i.e, that (lattice) universality holds for this model.

## Acknowledgements

R.V. thanks the "Centre de Supercomputació de Catalunya" for a fellowship and the Physics Department of Embry-Riddle Aeronautical University for kind hospitality during an extended visit. W.J. gratefully acknowledges a Heisenberg fellowship by the Deutsche Forschungsgemeinschaft. We further thank The Florida High Technology and Industry Council for partial funding under Contract FHTIC-15423ERAU, and SCRI for providing us with computer time on their cluster of fast RISC workstations.